\documentclass[aps,prl,twocolumn,showpacs,floatfix]{revtex4-1}
\usepackage{graphicx}
\usepackage{amssymb}
\usepackage{dcolumn}
\usepackage{bm}
\usepackage{dcolumn}

\def\bea{\begin{eqnarray}}
\def\eea{\end{eqnarray}}
\def \be{\begin{equation}}
\def \ee{\end{equation}}

\begin{document}
\title{Active tuning of synaptic patterns enhances immune discrimination}
\author{Milo\v{s} Kne\v{z}evi\'{c}}
\author{Shenshen Wang}
\email{shenshen@physics.ucla.edu}
\affiliation{Department of Physics and Astronomy, University of California Los Angeles, Los Angeles, CA 90095, USA}
\date{\today}

\begin{abstract}
Immune cells learn about their antigenic targets using tactile sense: during recognition, a highly organized yet dynamic motif, named immunological synapse, forms between immune cells and antigen-presenting cells (APCs). Via synapses, immune cells selectively extract recognized antigen from APCs by applying mechanical pulling forces generated by the contractile cytoskeleton.
Curiously, depending on its stage of development, a B lymphocyte exhibits distinct synaptic patterns and uses force at different strength and timing, which appear to strongly impact its capacity of distinguishing antigen affinities. However, the mechanism by which molecular binding affinity translates into the amount of antigen acquisition remains an unsolved puzzle.
We use a statistical-mechanical model to study how the experimentally observed synaptic architectures can originate from normal cytoskeletal forces coupled to lateral organization of mobile receptors, and show how this active regulation scheme, collective in nature, may provide a robust grading scheme that allows efficient and broad affinity discrimination essential for proper immune function.
\end{abstract}

\maketitle

Cell-cell communication in the adaptive immune system is a multi-channel process that involves mechanical interactions, biochemical signaling, and direct material fluxes~\cite{janeways, huse:17}.
Activation of B lymphocytes depends on productive binding of B cell receptors (BCRs) to antigen (Ag) displayed on the antigen presenting cells (APCs)~\cite{carrasco:07, phan:09, junt:07, qi:06, gonzalez:10}.
During recognition, a highly organized yet dynamic motif, termed immunological synapse, forms at the intercellular junction, through which B cells engage and extract Ag from the APCs by applying mechanical pulling forces generated by the contractile cytoskeleton to which BCRs anchor~\cite{huse:17, batista:01, tolar:13, tolar:14, tolar:17}.

\begin{figure*}[t]
  \includegraphics[width=2\columnwidth]{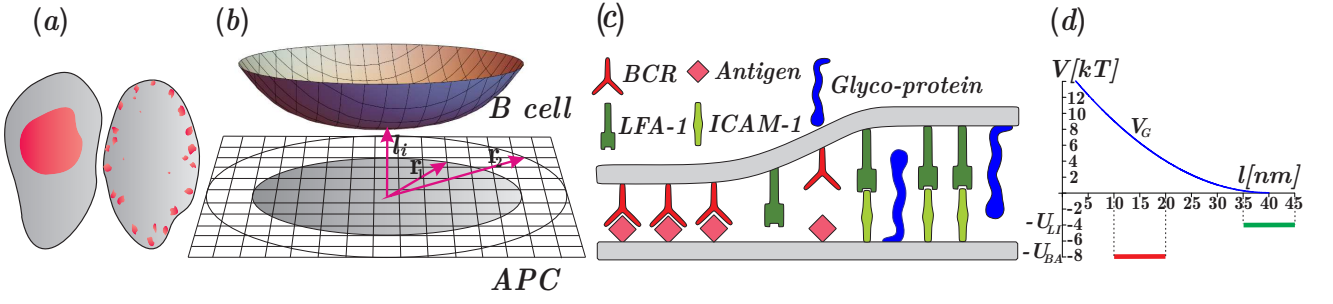}
  \caption{Model schematic. (a) Cartoon of immune synapses formed by a na\"{\i}ve or memory B cell (left) and a maturing B cell (right) with APCs. Gray: B cell-APC interface; red: BCR-bound Ag cluster(s). (b) 3D view of the initial profile of discretized membranes. The square patches have side length $a=70\,\mathrm{nm}$. Here $l_i$ measures membrane separation at patch $i$. Shaded area within the inner circle (radius $r_1$) indicates the contact zone; outer circle (radius $r_2$) encloses the simulation domain. (c) Side view of membrane proteins that are laterally mobile and contributing to normal interactions. (d) Local interactions between apposing membranes: short-range binding potentials of BCR and Ag, $V_{BA}$ (red) and
  of LFA-1 and ICAM-1, $V_{LI}$ (green); steric repulsion due to glyco-repellers, $V_{G}$ (blue).
  Affinity $U_{BA}$ is varied in the simulations while $U_{LI} = 4 \, k_BT$ is fixed.
  }
\label{model}
\end{figure*}

Very recent experiments~\cite{tolar:16}
have shown vividly that, at various developmental stages, a B cell exhibits markedly distinct synaptic architectures and uses pulling forces at different timing and magnitude:
Na\"{\i}ve (Ag-unexperienced) and memory (differentiated) B cells form a large Ag cluster in the synapse center (Fig.~\ref{model}a left) prior to application of force~\cite{tolar:13,tolar:14},
whereas maturing B cells extract Ag using small peripheral clusters (Fig.~\ref{model}a right) that co-localize with actin filaments and myosin-II motors~\cite{tolar:16}, suggesting that pulling forces apply on individual contacts during their formation.

Active research on membrane adhesion has yielded valuable insight into conformations and patterns~\cite{hammer:87, lipowsky:86, goldstein:88, lipowsky:91, membrane:95, weikl:02b, gordon:08, smith:06, smith:09, sackmann:14, schmidt:15} as well as nonequilibrium behaviors, e.g. enhanced shape fluctuations~\cite{prost:96, manneville:01, ramaswamy:00, gov:05, gov:06, rozycki:06} or lateral diffusion~\cite{granek:99}, patchiness under recycling~\cite{gheber:99, turner:05, gomez:13}, curvature-mediated remodeling~\cite{mcmahon:05, mishra:08}, and generation of interfacial forces~\cite{schwarz:13}. Yet, the questions as to how \emph{normal} forces affect the assembly of synapses and why different spatial patterns are created in developing immune cells remain open.




Considering functional needs, na\"{\i}ve and memory B cells
mount digital response via ``thresholding" behavior, i.e., they become activated and start dividing when bound with high-affinity Ag but stay resting otherwise. In contrast, maturing B cells undergo affinity maturation~\cite{victora:12}, an evolutionary process of mutation, competition and proliferation, that ultimately generates high-affinity antibodies (i.e. secreted BCRs). Effective operation of natural selection requires a ``grading scheme" that ranks B cells based on their affinity for encountered Ag~\cite{Batista:98, shih:02, victora:10, schwickert:11, victora:16}.
But an essential first step is poorly understood -- how molecular affinity of BCRs for Ag translates into the total amount of Ag that a B cell extracts from the APC. Given that the desired outcome is a \emph{gradual} dependence of Ag acquisition on affinity over a wide range, it would be intriguing to see how much discrimination can be achieved by synaptic contacts purely through the process of mechanical pulling.
Existing theories of immune synapses have assumed that cytoskeletal forces play a supporting role in reinforcing the contact pattern, without altering its nature; these models~\cite{chakraborty:01, weikl:02, chakraborty:03, weikl:04, dharan:17} indeed capture the complete phase separation between receptor-ligand complexes and bound adhesion molecules in na\"{\i}ve and mature cells~\cite{monks:98, dustin:98, dustin:99, dustin:07}.
However, none of them could explain the formation of persistent multifocal structures characteristic of maturing cells~\cite{hailman:02, richie:02, brossard:05, tolar:16}, in which aggregation of ligand-bound receptors and pulling on synaptic contacts may no longer be independent.




In this paper, we investigate the role of forces \emph{during} synaptic pattern formation.
Building upon a multi-component adhesive membrane model~\cite{weikl:04}, we incorporate mechanical forces that pull on BCR-Ag clusters.
This model recapitulates the experimentally observed variety of patterns and predicts new ones.
In essence,
a B cell can use normal pulling forces to actively control transitions between distinct patterns, by tuning the degree of phase separation. We further show that normal forces, when coupled to lateral organization of receptors, could enhance Ag affinity discrimination.

\textit{Model}.---
As in a typical \textit{ex vivo} experiment, a B cell is introduced to the proximity of a flat cell membrane, which mimics that of a stiff APC (Fig.~\ref{model}b).
Apposing membranes are discretized (with lattice constant $a$) and have two concentric circles inscribed into the lattices~\cite{bresenham:65} -- the inner circle (radius $r_1 = 45a$) encloses the contact region between the B-cell and APC membranes, while the outer circle (radius $r_2 = 100a$) delineates the simulation domain.
Membrane conformations are described by local membrane separation $l_i\geq0$ at the lattice sites $i$.
Five types of molecules (surveyed in Fig.~\ref{model}c) can occupy and hop between lattice patches (of size $a^2$) and mediate local adhesion or repulsion between membranes.
The B-cell membrane contains BCRs, integrin receptors LFA-1, and glyco-proteins, described by patch occupation numbers $n_i^{B}$, $n_i^{L}$ and $n_i^{Gb}$, respectively. Correspondingly, the APC membrane contains Ag, ICAM-1 (ligands of LFA-1) and glyco-proteins, with occupation numbers $n_i^{Ag}$, $n_i^{I}$ and $n_i^{Ga}$.
All molecules can diffuse across the simulation domain ($r\leq r_2$) but only interact in the contact zone ($r\leq r_1$).
The chosen patch size allows several molecules to occupy a single patch.

The overall configurational energy of the system in the contact region (Eq.~\ref{Ham}) consists of three contributions: the elastic energy $H_{el}$ of the membranes (Eq.~\ref{Hamel}), the interaction energy $H_{in}$ of receptors, ligands and glyco-repellers (Eq.~\ref{Hamin}), and the mechanical energy $H_{me}$ associated with normal pulling forces exerted on BCR-Ag bonds (Eq.~\ref{Hamf}):
\bea
\mathcal{H} (n, l; F) &=& \mathcal{H}_{el}(l) + \mathcal{H}_{in} (n, l) + \mathcal{H}_{me} (n; F), \label{Ham} \\
\mathcal{H}_{el} (l) &=& \sum_i \left [ \frac{\kappa}{2a^2} (\Delta_d l_i)^2 + \frac{\sigma}{2} (\nabla_d l_i)^2 \right ], \label{Hamel} \\
\mathcal{H}_{in} (n, l) &=& \sum_i [ \min(n_i^B, n_i^{Ag}) V_{BA}(l_i) \nonumber \\
+ \min(&n_i^L&, n_i^I) V_{LI}(l_i) +  (n_i^{Gb} + n_i^{Ga}) V_G(l_i) ], \label{Hamin} \\
\mathcal{H}_{me} (n; F) &=& \sum_j \sum_{i \in C_j} \min(n_i^B, n_i^{Ag}) F \Delta l_j \label{Hamf}.
\eea
The elastic deformation of the B-cell membrane is governed by the bending rigidity $\kappa$ and the lateral tension $\sigma$ (Eq.~\ref{Hamel}); we choose $\kappa = 12.25 \, k_BT$~\cite{seifert:95} and $\sigma = 0.1 \kappa/a^2$~\cite{simson:98}.
The Laplacian $\Delta_d$ and gradient $\nabla_d$ operators take a discretized form [SI].
In Eq.~\ref{Hamin}, the short-range binding potential for BCR and Ag is given by
$V_{BA}(l_i) = \{ - U_{BA}, \, 10 \, \textnormal{nm} \le l_i \le 20 \, \textnormal{nm}; \, 0, \, \textnormal{otherwise} \}$, where $U_{BA}$ is the binding affinity; see Fig.~\ref{model}d. Similarly,
$V_{LI}(l_i) = \{ - U_{LI}, \, 35 \, \textnormal{nm} \le l_i \le 45 \, \textnormal{nm}; \, 0, \, \textnormal{otherwise} \}$.
The term $\min(n_i^B, n_i^{Ag})$ denotes the minimum of the numbers of BCR and Ag molecules at patch $i$ and hence represents the number of interacting BCR-Ag pairs therein.
The repulsive potential due to glyco-repellers is given by $V_{G} = U_{G} ( l_i - l_{G})^2$, where $l_{G} = 40 \, \textnormal{nm}$, and $U_{G} = 10 \kappa/a^2$.
The key ingredient for generating a stable multifocal BCR-Ag pattern is the mechanical energy $H_{me}$.
We consider a simple setting that couples normal force to lateral patterning: pulling forces only apply to clusters, $\{C_j\}$, whose sizes (given as the total number of topologically connected patches occupied by at least one BCR-Ag bound pair) are bigger than a threshold size $n_t$; individual bonds in an above-threshold cluster $C_j$ are equally stressed by a constant force $F$ and subject to a common membrane displacement $\Delta l_j$.
In Eq.~\ref{Hamf}, the index $j$ runs through the above-threshold clusters, whereas the index $i$ scans over BCR-Ag bound patches in each of these clusters.
Such cluster-size-dependent normal force is motivated by the feedback between receptor clustering and pulling activity via BCR signaling~\cite{tolar:13, tolar:14}.
We propagate the system using Metropolis-Hastings~\cite{metropolis:53,hastings:70} Monte Carlo simulations [SI].

\begin{figure}[t]
  \includegraphics[width=1\columnwidth]{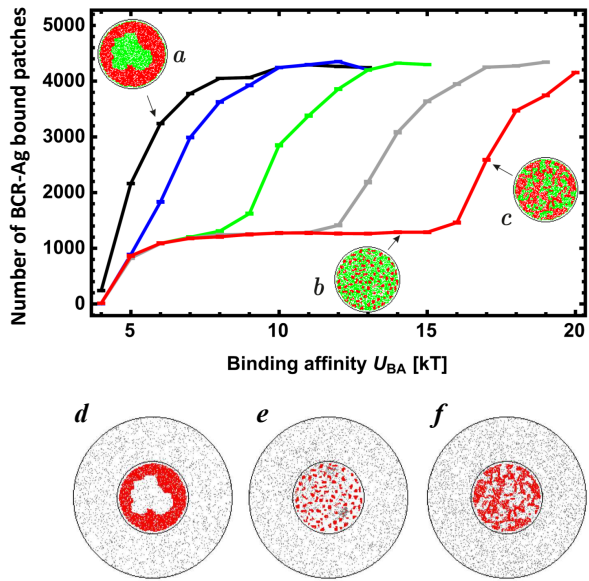}
  \caption{Affinity discrimination curves. Total number of apposing membrane patches bound by BCR-Ag pairs in the steady state is shown as a function of $U_{BA}$. Left to right: $Fl = 0, \, 6, \, 14, \, 22, \mathrm{and}\, 30 \, k_BT$. In the finite-force cases, normal pulling forces apply on clusters bigger than a threshold size $n_t = 20$. Insets (a) to (c) show synaptic patterns in various regimes; in the contact zone outlined in black, membrane patches bound by BCR-Ag pairs are shown in red, those bound by adhesion molecules in green. Panels (d) to (f) are wider views of (a) to (c), including the non-adhering region; red (gray) indicates membrane patches with BCR-bound (free) Ag.
  Lattice size is $L=200a$. The concentrations of BCR, LFA-1, ICAM-1 and glyco-repellers are $0.4/a^2$, while that of Ag is $0.3/a^2$.
  }
\label{summ}
\end{figure}

\textit{Results}.---Starting with uniformly distributed membrane proteins, as thermal undulations of membranes bring complementary receptors and ligands into proximity, synaptic patterns start to form and evolve.
Fig.~\ref{summ} shows Ag affinity discrimination curves (each for a given force $F$), which display the total number of apposing membrane patches that contain at least one bound BCR-Ag pair in the steady state, as a function of binding affinity $U_{BA}$. Equivalence of persistent BCR-Ag attachment to Ag extraction is valid when BCR-Ag bonds capable of withstanding disrupting forces are stronger than the Ag-APC association (see remarks in SI).

We first discuss the variety of steady-state patterns (Fig.~\ref{summ} insets; detailed features in Fig.~S1).
In the forceless scenario ($F=0$), we find two regimes, similar to those found earlier~\cite{weikl:04}.
For weak affinities, $U_{BA} \lesssim U_{LI}$, occasional BCR-Ag binding leads to transient Ag clusters dispersed in a background of bound adhesion molecules.
At higher affinities, complete phase separation occurs -- a single large cluster of BCR-Ag complexes (Fig.~\ref{summ}a) forms via continuous coarsening~\cite{bray:02} of smaller clusters (Movie a). Clustering and coarsening are driven by membrane elasticity which tends to minimize the line tension between Ag-rich and Ag-poor phases that differ in membrane separation.
If one waits long enough, the BCR-Ag ring (Fig.~\ref{summ}, a and d) eventually opens and a compact aggregate results.
This regime describes pattern formation in na\"{\i}ve and memory cell synapses prior to Ag extraction by force.

\begin{figure}[t]
  \includegraphics[width=1\columnwidth]{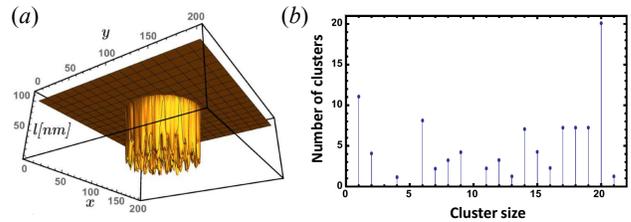}
  \caption{Arrested phase separation in model maturing B-cell synapses. (a) Membrane separation profile $l(x,y)$ in the contact zone. Here $x$ and $y$ are in unit of $a$. In the non-adhering region, membrane separation is kept at $l=100 \, \textnormal{nm}$.
  (b) Size distribution of BCR-Ag clusters shown in Fig.~2b. Steady state is reached in $2 \times 10^5$ Monte Carlo steps.
  $U_{BA} = 10 \, k_BT$, $n_t = 20$ and $Fl = 22 \, k_BT$.
  }
\label{dist}
\end{figure}

Under sufficiently strong pulling forces ($Fl\geq14\,k_BT$ in Fig.~\ref{summ}), synaptic patterning progresses through three regimes of Ag affinity.
(I) For modest affinities ($U_{BA} \lesssim U_{LI}$), similar to the no-force case, a lawn of bound adhesion complexes is punctuated by sparse BCR-Ag clusters (Movie b). Yet force-induced bond breakage further hinders the nucleation and growth of clusters, thus raising the affinity threshold for finite attachments and reducing the surface coverage of bound Ag.

(II) As the affinity increases, a new patterning regime emerges, visible as the (approximate) plateau in the affinity discrimination curve (e.g. the interval $U_{BA} \in [5,15] \, k_BT$ at $Fl=30 \, k_BT$, red curve in Fig.~\ref{summ}).
In this regime, small BCR-Ag clusters nucleate and begin to grow.
In the absence of pulling forces, these clusters would continue to coarsen and the system would proceed to complete phase separation. However, once the clusters reach the threshold size $n_t$, pulling forces set in and limit their growth in two ways: first, pulling increases the dissociation rates of individual BCR-Ag bonds~\cite{bell:78}, making it more likely that bound BCRs or Ag unbind and hop to a neighboring site, reducing the cluster size; second, puling acts on each of the above-threshold clusters as a whole (Eq.~\ref{Hamf}) and promotes rupture of entire clusters upon membrane displacements.
Opposing these bond-losing processes, protein influx from the reservoir in the non-adhering region $(r_1 < r \le r_2)$
supplies to the contact zone unbound pairs of BCRs and Ag that could either join existing clusters or form new ones. Once these processes strike a balance, phase separation is arrested -- clusters do not coarsen further (Fig.~\ref{summ}, b and e) and their sizes strongly peak at $n_t$ (Fig.~\ref{dist}b). The resulting multifocal pattern mirrors the one observed in maturing B cells~\cite{tolar:16}.
The balance between bond formation and dissociation is maintained at a dynamic equilibrium; while the cluster size distribution remains largely steady, constant material fluxes between clusters manifest as migrating clouds of free Ags in the contact zone (Movie c), where domains of dense unbound Ags (in gray, Fig.~\ref{summ}e) were just set free by cluster rupture and single-bond breakage.
Moreover, the B-cell membrane exhibits dynamic pod-like protrusions that concentrate BCRs at their contact sites with the APC (Fig.~\ref{dist}a), closely resembling structures formed \textit{in vivo}~\cite{kwak:17, jung:16}.

(III) 
Once the affinity is sufficiently high to overcome the disrupting effect of pulling forces, clusters larger than $n_t$ begin to appear in the pattern, and the plateau in the discrimination curve ends. This occurs at a force-dependent affinity value ${U}^*_{BA}(Fl)$ (e.g., ${U}^*_{BA} \approx 15 \, k_BT$ for $Fl = 30 \, k_BT$)
and marks the onset of a distinct regime where merging clusters percolate through the contact zone (Fig.~\ref{summ}, c and f). Importantly, this percolating structure shows no tendency toward coalescing into one compact aggregate (Movie d), indicating a persistent influence of pulling forces that act to halt further coarsening and maintain a ramified morphology.
Only when the affinity is considerably larger than ${U}^*_{BA}(Fl)$ does complete phase separation take place; the amount of long-lived BCR-Ag bonds tends to the no-force value.
The shift in the coarsening transition with increasing force can be described by a phenomenological model [SI].

\begin{figure}[t]
  \includegraphics[width=0.85\columnwidth]{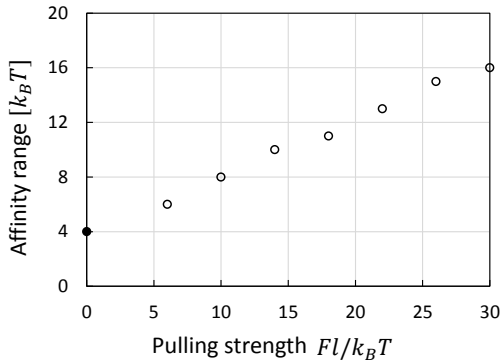}
  \caption{Capacity of affinity discrimination. The range of discernible affinity is shown as a function of the strength of pulling force. The filled symbol indicates the force-free value, whereas open symbols correspond to finite forces. $n_t=20$.
  }
\label{capacity}
\end{figure}

As spatial patterns transition from transient clusters to multifocal contacts to a percolating network, the area coverage of Ag that remain bound to BCRs also varies with Ag affinity. In contrast to the no-force situation, where Ag coverage rapidly saturates at modest affinity (Fig.~\ref{summ}, black curve), strong pulling forces on above-threshold clusters not only extend the discernible limit to higher affinities, but lead to a gradual monotonic dependence indicating discrimination of small differences over a wide range (Fig.~\ref{summ}, green, gray and red curves).
This ``grading scheme" stems from affinity-dependent abundance and size of Ag clusters (Fig.~S2). In regime I, as affinity increases, a greater number of clusters form and grow to larger sizes (all below $n_t$ though) and hence a rapid increase in Ag coverage. In regime II, force limits the cluster size and yet a gradual increase in cluster abundance accounts for the slight increase in Ag coverage. In regime III, clusters rapidly exceed $n_t$; smaller structures merge into larger ones which are more resistant to pulling forces. Therefore, the rise in Ag coverage results from an increasing inability of force to disrupt BCR-Ag bonds.

Finally, we find that active testing of bond strength using pulling force that depends on cluster size significantly enhances discrimination capacity and efficiency.
Discrimination capacity, measured by the range of discernible affinity (above the minimum affinity for finite attachment and below the maximum affinity prior to saturation), increases with the strength of pulling (Fig.~\ref{capacity}), owing mainly to an extended multifocal regime [SI].
Furthermore, when forces apply (Fig.~S3b), broad discrimination can be realized within minutes, comparable to the lifetime of B cell synapses \textit{in vivo}~\cite{tolar:14}. Without pulling (Fig.~S3a), it takes an hour to distinguish the same set of affinities to a lower quality.

\textit{Discussion}.---Each maturing B cell performs numerous parallel pulling experiments to test the quality of Ag binding.
We suggest that application of normal mechanical stresses not merely affects bond dissociation \emph{after} synaptic patterns form and thereby regulates Ag extraction, but directly participates in the patterning process so that Ag coverage responds to the competition of pulling and binding in approach to the steady state. Further, we show that mechanical pulling during synapse formation could enhance affinity discrimination, not only through increasing the off rate of weak bonds thus thresholding for high-affinity Ag, but via grading the Ag that pass the threshold by tuning the number and size of Ag clusters over a broader range of affinities. Therefore, distinct from schemes (e.g. conformational proofreading~\cite{savir:07}) that enhance the specificity of one-on-one binding reactions, broad discrimination of maturing B cells is achieved through \emph{collective} effects of cluster-size-dependent pulling forces that induce and modulate spatial patterns of receptors in multimeric binding responses.

We show that coupling of normal forces to lateral organization of membrane proteins can lead to steady-state clusters that are \emph{intermediate} in size. This result agrees with experimental observations of multifocal patterns formed in maturing B-cell synapses and in immature T cells during thymic selection, in stark contrast to the large domains observed in na\"{\i}ve or memory cell synapses. This distinct patterning regime reveals a recycling mechanism: the Ag deposition/removal process (due to bond formation/rupture) effectively arrests the spinodal decomposition induced by membrane-mediated lateral attractive interactions.
Arrested phase separation is robust to change in the threshold cluster size $n_t$ as long as $n_t>1$ [SI]; a finite yet modest $n_t$ promotes persistent discrimination over a wide affinity range, while too large or too small $n_t$ desensitizes patterning response to affinity change. Multifocal patterns also form in a load-sharing setting~\cite{erdmann:04, erdmann:04b} but at a higher energy cost [SI]. Our model also predicts best discrimination at intermediate Ag concentrations [SI].


In sum, a primary result of this work is that mechanical tuning of synaptic patterns via normal forces exerted by the cytoskeleton could represent an important cellular strategy of efficient immune discrimination. This proposal may be tested by monitoring spatiotemporal dynamics and force usage during synapse formation, as distinct patterning regimes are traversed experimentally, either by varying the strength of force (e.g. by modifying the activity of myosin II motors) at fixed Ag affinity, or by changing affinity without affecting the force (e.g. by using different types of Ag).
Our results suggest novel targets of manipulation for high-quality antibodies, and provide insights into biomimetic design for selective recognition and controlled adhesion.



\textit{Acknowledgments}.--- We thank Robijn Bruinsma for fruitful discussions. This work has been supported by UCLA startup grant (S.W.).

\bibliography{Ref}


\end{document}